\documentclass{ws-procs9x6}

\begin{document}

\title{Cooling delay for protoquark stars due to neutrino trapping
\footnote{\uppercase{C}onference \uppercase{P}roceedings of the 
\uppercase{KIAS-APCTP} \uppercase{I}nternational
\uppercase{S}ymposium in \uppercase{A}stro-\uppercase{H}adron 
\uppercase{P}hysics; \uppercase{C}ompact \uppercase{S}tars: 
\uppercase{Q}uest for \uppercase{N}ew \uppercase{S}tates of 
\uppercase{D}ense \uppercase{M}atter}}
\author{J.~Berdermann
}
\address{Fachbereich Physik, Universit\"at Rostock,
        D--18051 Rostock, Germany\\
         E-mail: jens@darss.mpg.uni-rostock.de}
\author{ D.~Blaschke}
\address{Fachbereich Physik, Universit\"at Rostock,
        D--18051 Rostock, Germany\\
        Bogoliubov Laboratory for Theoretical Physics, JINR Dubna,
        141980 Dubna, Russia
        }
\author{H.~Grigorian
}
\address{Fachbereich Physik, Universit\"at Rostock,
        D--18051 Rostock, Germany\\
        Department of Physics, Yerevan State University,
        375025 Yerevan, Armenia
        }

\maketitle

\abstracts{The influence of neutrino trapping (NT) on the early cooling evolution of
hot proto quark stars (PQS) with initial temperatures in the range $T\sim 40~$ MeV
is studied. Within a simplified model for the neutrino transport it is shown
that the time for reaching neutrino opacity temperature of
$T_{opac}\sim 1$ MeV is about 10 sec.
This is an order of magnitude larger than
without NT and of the same order as the duration long gamma ray bursts.
 }


\section{Introduction}
\label{sec:intro}
Gamma ray bursts (GRBs) are among  the most intriguing phenomena
in the Universe, see \cite{Piran} and references therein.
If the energy is emitted isotropically, the measured
energy release is of the order of $10^{53}\div 10^{54}$~erg
and it is a puzzle to explain the engine of a GRB.
However, there is now a compelling evidence that
the gamma ray emission is not isotropic, but displays a jet-like geometry.
When the emission is collimated within a narrow beam a  smaller energy,
of the order of $10^{52}$~erg, would be sufficient for the GRBs \cite{Frayl}
but their sources are not yet understood.
There is growing evidence for a connection of GRBs to supernovae
now from emission features in some GRB afterglows, e.g.
GRB 990707 \cite{Amati}, GRB 991216 \cite{Piro}, GRB 000214 \cite{Antonelli}
and, most recently GRB 030329 \cite{Fynbo}.
As in the realm of a supernova explosion a compact star is likely to be born,
it has been conjectured (see \cite{Berezhiani} and references therein)
that a phase transition from hadronic to deconfined quark matter might
power the GRB. However, although the energy release might be in the
right order of magnitude, the collapse timescale is too short ($\sim $
several ms)
to explain long GRBs with a duration of several tens of seconds.
Recently, it has been suggested \cite{Berezhiani} that the deconfinement
transition in a compact star might be an example for a nucleation process of
quark matter droplets which is a quantum tunneling process between metastable
states, with a sufficient
delay, depending on the surface tension of the quark matter droplet.
This approach has been criticized \cite{DNV} since during the
supernova collapse the protoneutron star can be heated up to
temperatures of the order of the Fermi energy
$T\sim \varepsilon_F \approx 30 \div 40$ MeV so that the thermal fluctuations
would dominate over quantum ones and make the phase transition
sufficiently fast without the delay claimed in Ref. \cite{Berezhiani}.

In the present contribution we consider the cooling evolution of a hot
protoneutron star above the neutrino opacity temperature $T_{opac}\sim 1$ MeV
\cite{Reddy}, so that
the neutrino mean free path is by orders of magnitude smaller than the
size of the star.
As it has recently been estimated \cite{Aguilera} and also reported at this
conference \cite{Agui_Korea}, the neutrino untrapping transition
occuring when the star cools below $T_{opac}$ might serve as an engine of
a GRB.
The question to be considered here is whether the energy release by
neutrino emission
can be sufficiently delayed due to neutrino trapping so that the typical
duration of a long GRB could be explained.

\begin{figure}[th]
\psfig{figure=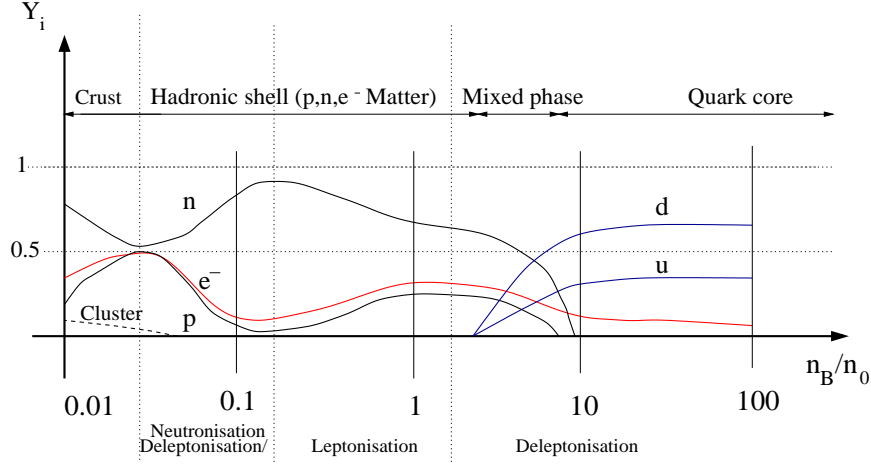,width=\textwidth,angle=0}
\caption{Schematic representation of the evolution
 of the composition of compact star matter with increasing
 density. $Y_i=n_i/n_B$ is the fraction of particles of species i per
baryon and $n_B/n_0$ is the baryon density in units of the nuclear
saturation density $(n_0 = 0.16~ {\rm fm}^{-3})$.
\label{Graph15e}}
\end{figure}

\section{Early cooling evolution}
\label{sec:cool}

During the collapse of the progenitor star the density increases in
its very interior from the densities of the iron core to those at and
above the deconfinement transition.
For this huge interval of densities, part of which we display in
Fig. 1,
matter can undergo several phase transitions.
During these processes, like leptonization  and deleptonization,
neutrinos and antineutrinos are produced in
weak interactions which
eventually could also be trapped during the early stages of the
evolution when the temperatures are expected to be much higher than
 $T_{opac}$.
The cooling process for such a PQS is investigated in this contribution,
where we use a simplified model of a homogeneous PQS structure and an
approximate global thermal evolution.
If the star cools below $T_{opac}$ by surface-radiation, neutrinos
could escape in a sudden outburst.
For brevity, saying neutrino means both neutrino and antineutrino.

The loss of energy in a homogeneous system due to emission is given by
\begin{equation} \label{f6.1}
\frac{dU(T)}{dt} = -\epsilon_{\nu}(T) \cdot V~.
\end{equation}
Here the photon emissivity has been neglected since neutrinos dominate
the cooling evolution of a PQS for temperatures well above $10^6$ K.
For the cooling behaviour of a star from a given initial-temperature $T_i$
to a final temperature  $T_f$ with the luminosity
$L(T) = -\epsilon_{\nu}(T) \cdot V$ follows
\begin{equation} \label{f6.2}
 \triangle t = - \int\limits_{T_i}^{T_f} \frac{C_v(T) \cdot dT}{L(T)} ~.
\end{equation}
The luminosity $L(T) = L_{\bar{\nu}}^{V}(T)+L_{\bar{\nu}}^{O}(T)$
is explained more in detail in Section \ref{sec:lumi} and the specific
heat $C_v(T)$ can be derived from the thermal energy $U(T)$  for
relativistic quark matter
\cite{Shapiro}
\begin{equation} \label{f6.3}
C_v(T) = \frac{dU}{dT} = 1.8\times 10^{48}
\biggl(\frac{M}{M_{\odot}}\biggl)\cdot
\biggl(\frac{n_B}{n_0}\biggl)^{-1/3} T_9
\hspace{0.2cm} {\rm erg}.
\end{equation}
Here $M$ is the mass of the star in units of the
solar one and $n_B$ is the baryon density of the star for
a homogeneous mass distribution in units of the nuclear saturation
density $n_0=0.16~{\rm fm^{-3}}$.
For the temperature we use the standard notation $T_9 = T/10^9 ~{\rm K}$.

\begin{figure}[thb]
\vspace{-1cm}
\psfig{figure=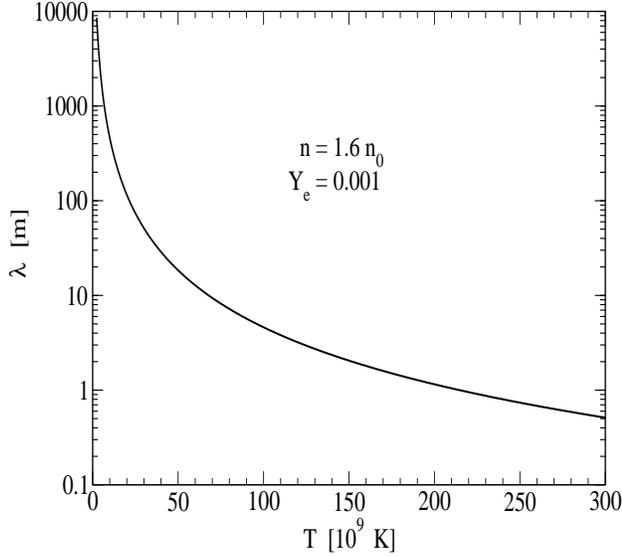,height=9cm,width=9cm,angle=-90}
\caption{The neutrino MFP as a function of the
temperature for given baryon density and electron fraction.
\label{b6.1}}
\end{figure}

\begin{figure}[htb]
\vspace{-1cm}
\psfig{figure=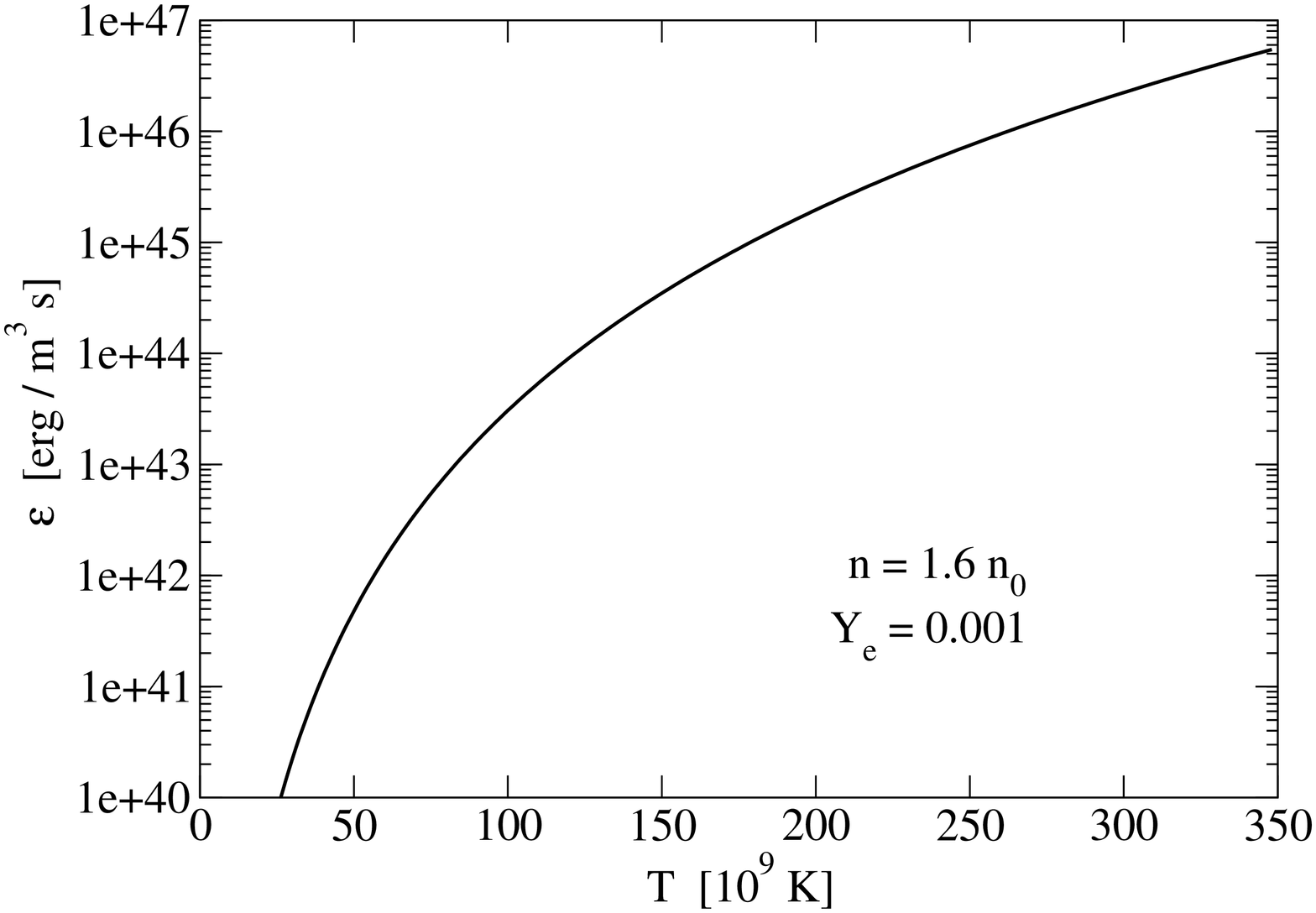,height=9cm,width=9cm,angle=-90}
\caption{Temperature dependence of the neutrino
emissivity for the URCA process for given baryon density and electron
fraction.
\label{b6.2}}
\end{figure}

The emissivity for the direct
URCA process in normal quark matter is given by
\begin{equation} \label{f6.6}
\epsilon^{URCA}_{\nu}(T) \simeq \frac{914}{315}\pi^2 G_F^2 {\rm cos}^2
  \theta_c \alpha_s~ {n_B}(3Y_d~ Y_u~Y_e)^\frac13~T^6,
\end{equation}
where $G_F$ is the Fermi constant of the weak interaction, $\alpha_s$ is the
strong coupling constant and $\theta_c$ the Cabbibo angle.
We have introduced the fractions of the particle species $i$
as  $Y_i = n_i/n_B$ and the baryon density in terms of up and down quark 
densities is $n_B = (n_u+n_d)/3$.
The neutrino mean free path (MFP) has the expression \cite{Iwamoto}
\begin{eqnarray} \label{f6.7}
\lambda(T)&=&\frac{(6\pi)^\frac13}{12G_F^2 {\rm cos}^2\theta_c}
 \frac{Y_\nu^\frac13}{Y_{u}^{\frac23}Y_e}{n_B}^{-\frac43}
  \left[1+\frac12\left(\frac{3Y_e}{Y_u}\right)^\frac13+\frac{1}{10}
\left(\frac{3Y_e}{Y_u}\right)^\frac23\right]^{-1} \nonumber\\
 &&\times[(E_{\nu}-\mu_{\nu})^2+(\pi T)^2]^{-1}.
\end{eqnarray}
Using the PQS matter constraints of  charge neutrality
$\frac{2}{3}n_u-\frac{1}{3}n_d-n_e=0$ and $\beta$-equilibrium for the case 
of trapped neutrinos $\mu_{\nu} = \mu_u + \mu_e -\mu_d$,
we can express all particle fractions $Y_i$ via $Y_e$. 
In our model calculation we choose $Y_e = 0.001$ and approximate 
$E_{\nu} \simeq T$. 
For the above choice of parameters, the temperature dependence of the MFP
is shown in Fig. \ref{b6.1}  and the emissivity in Fig. \ref{b6.2},
respectively.

\section{Emissivity and luminosity of PQS}
\label{sec:lumi}
\begin{figure}[ht]
\psfig{figure=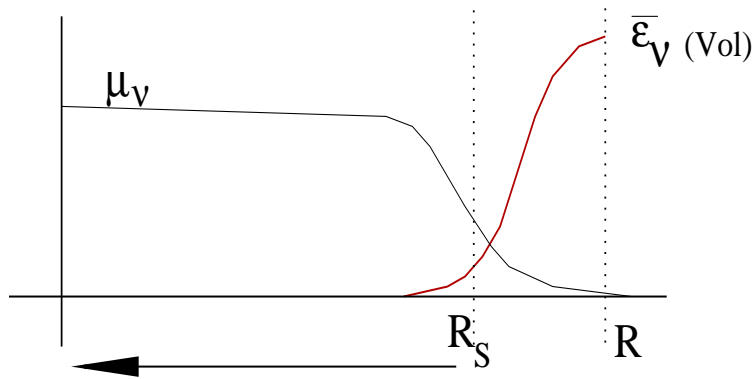,height=5cm,width=10cm,angle=0}
\caption{Evolution of the neutrinosphere.
\label{Graph3}}
\end{figure}

The emissivity $\epsilon_{\nu}^{URCA}$  for the direct URCA-process in
quark-matter Eq. (\ref{f6.6})  has to be modified for temperatures,
at which the star is opaque to neutrinos.
Due to the trapping of the neutrinos their emissivity is modified by
a factor which takes into account the probability that the neutrino
created at a distance $r$ from the center can leave the star in the
direction given by the angle $\alpha$.
The effective emissivity is given by a product of the emissivity for
the direct URCA process and an exponential suppression factor
\begin{equation} \label{f6.12}
\bar{\epsilon}_{\nu}(r,\alpha;T) =
{\exp}\left[-l(r,\alpha)/\lambda(T)\right]\cdot
\epsilon^{URCA}_{\nu}(T) ~,
\end{equation}
where $l(r,\alpha)$ is the distance from the neutrino creation point
to the star surface, which for a spherically symmetric star with
radius $R$ is given by
\begin{equation} \label{f6.13}
l(r,\alpha)=\sqrt{R^2-r^2 {\sin}^2 \alpha}-r~ {\cos}\alpha.
\end{equation}
We average over all possible neutrino directions
\begin{equation} \label{f6.14}
\bar{\epsilon}_{\nu}(r;T)= \frac{1}{\pi} \int\limits_{0}^{\pi}d\alpha~
\bar{\epsilon}_{\nu}(r,\alpha;T)~,
\end{equation}
and integrate over all distances $r$ up to the star radius $R$ in order to
obtain the total luminosity for neutrino emission from the star volume
\begin{eqnarray} \label{f6.15}
L_{\nu}^V(T) &=& 4 \pi \int\limits_{0}^{R} dr~ r^2
\cdot\bar{\epsilon}_{\nu}(r;T)~.
\end{eqnarray}
As long as the temperatures are high enough, $T \stackrel{>}{\sim}1$ MeV,
there is a spherical inner star region, from where practically no
neutrinos escape so that their number is quasi conserved and can be
defined by a finite chemical potential $\mu_\nu$.
This region extends up to a distance $R_S$ from the center.
The radius $R_S$ of the neutrinosphere
is a function of the temperature and
moves towards the center during the cooling evolution, see Fig. \ref{Graph3}.
In calculating the bulk neutrino luminosity of a PQS during the
trapping era, we can restrict the integration in Eq. (\ref{f6.15})
to the region between the neutrinosphere and the surface of the star.
Besides the bulk luminosity we have additional neutrino radiation just
from the neutrinosphere which can be taken into account by the
generalized blackbody radiation formula
\begin{equation} \label{f6.16}
L_{\bar{\nu}}^{O}(T)=4\pi R_S^2 \cdot
\left(\frac{\mu_\nu^4}{4\pi^2}+
\frac{\mu_\nu^2~T^2}{2}+\frac{7 \pi^2~T^4}{60}\right)~,
\end{equation}
which we denote as inner surface luminosity.
From the evolution of $R_S$ with time we can characterize the
untrapping transition of neutrinos as a burst-type phenomenon or a
smooth fading.
\begin{figure}[ht]
\vspace{-1cm}
\psfig{figure=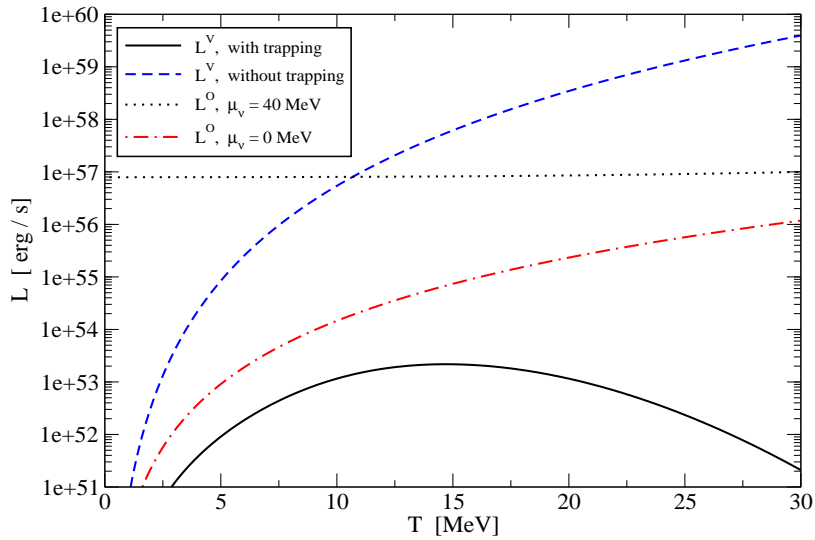,height=9 cm,width=12cm,angle=-90}
\caption{Temperature dependence of the bulk and inner surface
neutrino luminosities.
\label{b6.8}}
\end{figure}
In Fig. \ref{b6.8} we compare the temperature dependence of the
bulk and inner surface luminosities.
These results show that the trapping of the neutrinos changes the
luminosities by about 6 orders of magnitude within the trapping
regime.
Moreover, for the trapping case, in contrast to the untrapped bulk
emission, the bulk luminosity has a maximum at $10^{53}$ erg/s for
temperatures around 15 MeV.
Due to this particuliar behavior of the bulk luminosity in the
neutrino trapping regime we expect that for initial temperatures much
higher than $30$ MeV there is not only a quantitative change in the
cooling evolution but rather a qualitative change in the temporal
evolution of the energy release.

\begin{figure}[th]
\vspace{-1cm}
\psfig{figure=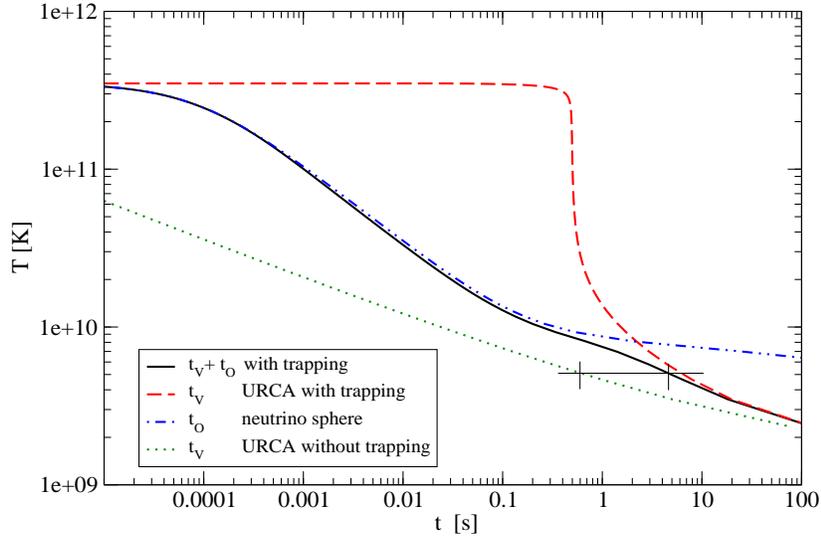,height=9cm,width=12cm,angle=-90}
\caption{{Cooling behavior of a protoneutron star with initial
    temperature $T_i=30$ MeV and $R=12$ km.}
\label{b6.9}}
\end{figure}

\section{Results}
\label{sec:result}
In Fig. \ref{b6.9} we show the cooling curves of a PQS for an
initial temperature of $T_i= 30$ MeV and a star radius
of $R=12$ km. The solid line corresponds to a cooler in the trapping
regime including all effects discussed above.
The other curves correspond to limiting cases, in particular to the
case without trapping (dotted line), the case with trapping but
without surface emission (dashed line) and the emission from the inner
sphere only (dash-dotted line).
The comparison shows that at the vicinity of the total untrapping
regime for $T=0.6$ MeV the time delay of the cooling with trapping
amounts to a factor of ten.
If one will neglect the inner surface emission of the neutrinos, then
the neutrino release will not only be delayed but also occur within a
sudden burst.

\section{Outlook}
\label{sec:out}
The investigation of the effects of neutrino trapping on PQS evolution
is only in its beginning. We have outlined a simple model for the study
of spherical neutrino emission from a hot PQS with and without
trapping.
Detailed investigations have to be conducted in order to make firm
conclusions whether the hot neutrino trapping scenario could shed
light into the mysteries of GRBs and the GRB supernova connection.
We underline two observations made in this report:
\begin{itemize}
\item The main thermal energy of the star is
released in a time interval, which is of the order of the duration
of long GRB's ($\sim 10$ s).
\item The amount of energy which can be released in the cooling of a
homogeneous PQS depends much on the initial temperature which is
unknown and might for the PQS scenario be up to one order of magnitude
larger than for the canonical protoneutron star scenario.
\end{itemize}
This latter observation can have significant implications for the
early cooling evolution
after a supernova collapse due to the strong temperature dependence of
the dominating URCA process on the one hand and the large effective
absorption of bulk emissivity on the other.
Note that we have omitted here the possible effects of finite thermal
conductivity which could make the separation of the neutrinosphere
from the bulk emission zone even stronger and enhance the eventual
temporal structure in the cooling evolution.
First estimates show that the neutrino release can occur within a
burst \cite{Aguilera} and provided the conversion process to gamma
rays is effective enough \cite{Haensel:um}, an interesting PQS-GRB
scenario emerges
which can include a beaming mechanism due to the formation of a vortex
lattice in a strongly magnetized superconducting PQS \cite{BBGHV}.


\section*{Acknowledgments}
J.B. and D.B. thank the organizers of the KIAS-APCTP International
Symposium on Astro-Hadron Physics of
Compact Stars and DAAD-HOST programm D/03/31497 for the financial support
of their conference participation and a study visit to Pusan University.
The authors thank the colleagues for their interest in this work and useful
discussions during the conference.
H.G. acknowledges the support of the Virtual Institute VH-VI-041 of
the Helmholtz Association for ``Dense hadronic Matter and QCD Phase
Transition''.

\end{document}